\documentclass[aps,prl,twocolumn,showpacs,superscriptaddress]{revtex4}
\usepackage{color}
\usepackage{graphicx}
\usepackage{amsmath}
\usepackage{amssymb}
\usepackage{hyperref}
\bibstyle{apsrev.bib}

\begin{document}

\title{Calculation of transport coefficient profiles in modulation experiments as an inverse
problem }

\author{D.F. Escande}

\address{UMR 6633 CNRS-Universit\'{e} de Provence, Marseille, France}

\address{Consorzio RFX, Associazione EURATOM-ENEA sulla fusione,
Padova, Italy}

\author{F. Sattin}

\address{Consorzio RFX, Associazione EURATOM-ENEA sulla fusione,
Padova, Italy}

\begin{abstract}
The calculation of transport profiles from experimental measurements belongs in the category of
inverse problems which are known to come with issues of ill-conditioning or singularity. A
reformulation of the calculation, the matricial approach,  is proposed for periodically modulated experiments, within the context of the standard
advection-diffusion model where these issues are related to the vanishing of the determinant
of  a 2x2 matrix. This sheds light on the accuracy of calculations with
transport codes, and provides a path for a more precise assessment of the profiles and of the
related uncertainty.

\end{abstract}


\maketitle

Inverse problems, i. e. going from data to model parameters, are ubiquitous in science and engineering. They are
known to come with issues of ill-conditioning or singularity, which make difficult relating the
accuracy of the computed model parameters to the errors in the input data. Inferring transport
properties of magnetically confined plasmas from perturbative experiments \cite{ryter} belongs in
this category. This issue is crucial to both the theoretical understanding of transport processes and
the practical control of fusion plasmas. Classical reconstructions of transport coefficients rely on
standard transport codes, which ignore the possible ill-posedness of the problem by employing \textit{ad hoc}
regularizations. To the best of our knowledge, so far the only authors that took it into account
were Andreev and Kasyanova \cite{andreev}, who provided a detailed analysis of the
uncertainties present in the case of a localized impulsive source. However, as shown later,
experimentalists happened unwittingly to avoid ill-conditioning and singularity by trying naturally
to separate the domain where transport is measured from that where sources are present.
Unfortunately this separation is almost impossible for measuring the transport of angular
momentum, and only partially possible for heat transport. It is therefore most important to warn
the users of transport codes, and to benefit from an alternative safer approach. For the first time
this paper proposes such an alternative. The corresponding
technique is elementary, quite general, and requires much less numerical calculations than with
transport codes. It may be applied to any experiment aiming at the calculation of transport
coefficients, also in media other than plasmas. Its only limitations are: (i) it works in effectively one-dimensional geometries; (ii) applies to perturbative regimes, i.e., where relevant equations may be linearized around unperturbed equilibria; (iii) the forcing term causing the perturbation must be periodic in time.\\
The standard procedure for inferring transport coefficients is through solving the advection-
diffusion  for the generic quantity $\zeta(r,t)$ (which may stand for the perturbation to particle density, temperature, momentum,...):
\begin{equation}
\begin{array}{l}
\partial_t \zeta  =  - \nabla  \cdot \Gamma (\zeta ) + S \\
\Gamma  =  - \chi \nabla \zeta  + V\zeta
\end{array}
\label{eq:fpe}
\end{equation}
In (\ref{eq:fpe}) $\chi, V$ are functions taken out of a set of trial profiles guessed a-priori,
usually simple analytical expressions containing a certain number of free coefficients which are
given explicit values by minimizing under some norm the gap between the experimental $\zeta$
profiles and the reconstructed ones. The regularization built in this approach is straightforward: it
consists in choosing well-behaved trial functions for $\chi, V$.\\
In this work we  address the problem of solving Eq. (\ref{eq:fpe}) as an inverse problem in the
case of periodically modulated perturbations. We  present a simple algebraic method to recover
\textit{exact} solutions; that is, $\chi, V$, may be exactly and directly (not by means of iterative
procedures) obtained from the smoothed data without the recourse to any adjustable trial
function and minimization procedure. We  identify under which conditions the problem is well-
or ill-posed, i.e., the solution $(\chi, V)$ does sensitively depend from the input data. We  discuss
how error bars in the raw measurements are propagated to final estimates of $(\chi, V)$. This is a
delicate issue when Eq. (\ref{eq:fpe}) is addressed as a least-squares problem using smooth trial
functions for $(\chi, V)$: from the one hand, they prevent any oscillation to blow up, thereby
constraining the apparent error bars to values smaller than ought be.  On the other hand the
opposite limit is also possible: if trial functions are chosen that cannot match the true $(\chi, V)$
profiles, the minimization procedure necessarily converges towards just a local minimum that is
far away from the true global minimum. In this case, the error on the final $(\chi, V)$ values  is
larger than expected from the raw data alone. Finally, we point out how data from different
experiments can be combined to reduce the effective margin of error and thus yield a more
accurate final estimate of $(\chi, V)$. \\
We  consider a case with cylindrical symmetry and a purely sinusoidal forcing term. Calculations
are easier using complex notation, thus the temporal behaviour is factored out in the term  $\exp(-i
\omega t) $ and Eq. (\ref{eq:fpe}) writes
\begin{equation}
- i \omega \zeta =  r^{-1} {\partial_r} \left[ r \left( \chi \partial_r \zeta - V \zeta \right) \right] + S
\label{eq:fpew}
\end{equation}
After rearranging, integrating over the radial coordinates and using the boundary condition
$\Gamma(r = 0) = 0$ we get
\begin{equation}
\chi \partial_r \zeta - V \zeta  = - r^{-1} \int_0^r \, dz \; z \left(i \omega \zeta + S \right)
\end{equation}
The previous equation is reduced to an algebraic system of two real equations by expressing the
signal in terms of amplitude and phase,
$\zeta = A e^{i \phi}$, and $S = S_r + i S_i$:
\begin{equation}
\begin{array}{l}
 {\bf{M}} \cdot {\bf{Y}} = {\bf{\Gamma }} \\
 {\bf{Y}} = \left( {\begin{array}{*{20}{c}}
   \chi   \\
   V  \\
\end{array}} \right),\quad {\bf{M}} = \left[ {\begin{array}{*{20}{c}}
   { - A'\cos \phi  + A\phi '\sin \phi } & {A\cos \phi }  \\
   { - A'\sin \phi  - A\phi '\cos \phi } & {A\sin \phi }  \\
\end{array}} \right], \\
\quad {\bf{\Gamma }} = \left[ {\begin{array}{*{20}{c}}
   {\frac{1}{r}\int\limits_0^r {dz\,z\left( {{S_r}(z) - \omega A(z)\sin \phi (z)} \right)} }  \\
   {\frac{1}{r}\int\limits_0^r {dz\,z\left( {{S_i}(z) + \omega A(z)\cos \phi (z)} \right)} }  \\
\end{array}} \right] \\
 \end{array}
\label{eq:emme}
\end{equation}
where the primes stand for differentiation with respect to $r$. Provided ${\bf \Gamma}(r)$ is
known, ${\bold Y}(r)$ can be computed by inverting the matrix ${\bold M}(r)$. For this reason,
in the following we  refer to this method as the matricial approach.
It should be mentioned that considering just time-periodic quantities in (\ref{eq:fpe}) leads to a
considerable simplification: this reduces Eq. (\ref{eq:fpe}) from a partial differential equation to an
ordinary differential equation (\ref{eq:fpew}). Several efforts were devoted to solve
Eq. (\ref{eq:fpew}) using simpler techniques than its numerical integration, possibly
even analytical methods \cite{jacchia}. \\
We now solve Eq. (\ref{eq:emme}) for ${\bold Y}(r)$ in terms of the eigenvalues  $\lambda(r)$ and
eigenvectors ${\bold E}(r)$ of the matrix ${\bold M}(r)$	
\begin{equation}
{\bf{Y}} = {y_0}{{\bf{E}}_0} + {y_1}{{\bf{E}}_1},\,\;{y_j} = \lambda _j^{-1}
{\bf{\Gamma }} \cdot {{\bf{E}}_j}  \quad (j = 0,1)
\label{eq:y}
\end{equation}
A solution exists wherever the matrix ${\bold M}$ is invertible, i. e. where none of the
$\lambda_j$'s vanishes, which would imply
\begin{equation}
 \det(M) = A^2 \phi' = 0
\label{eq:det}
\end{equation}
Excluding exceptional cases where $A = 0$, the singular points are those where $\phi' = 0$.
The origin is one such point, but there ${\bf \Gamma} = 0$, too; thus ultimately $r = 0$ is a
regular point. Inspection of literature
(see e.g. \cite{salmi,mantica08,mantica05,ryter00,lopez,mantica02}) shows instead that the
condition $\phi' = 0$ is usually fulfilled
close to the location of the source. Qualitatively, this can be justified by noticing that, close to the source, the
dynamics of $\zeta$  is ruled in Eq. (\ref{eq:fpew}) more by the source than by transport if
$\omega$ is large enough, hence $ \zeta \approx i S \omega^{-1}$ and $S' = 0 $ implies $ \phi' = 0$.
\\
Physically, transport coefficients are defined even at the singular points $r = r_s$. Throughout this work we conventionally choose $\lambda_0$ to be the eigenvalue that vanishes at $r_s$. Accordingly, the actual value of $\zeta$ produced by the source $S$ must align the flux ${\bf \Gamma}$ exactly along the ${\bf E}_1$ eigenvector:
\begin{equation}
{\bf{M}} \cdot {\bf{Y}} = {\lambda _1}{y_1}{{\bf{E}}_1},\quad {\bf{\Gamma }} =
{\gamma _1}{{\bf{E}}_1},\quad (r = r_s)
\end{equation}

Due to unavoidable uncertainties arising in the experiment, the estimated $\bf \Gamma$  takes on a component
along ${\bf E}_0$ too, thus making the inversion impossible. However we may solve for the
component of $\bf Y$ along ${\bf E}_1$, whereas the component along  ${\bf E}_0$ remains
completely unknown: ${\bf Y}(r_s)$ belongs to a straight line parallel to ${\bf E}_0$ in the plane $(\chi,V)$.

We present now a comprehensive discussion, including both regular and singular points. We label
with the subscript "m" the quantities measured by the experiment: ${\bf M}_m, \zeta_m,
{\bf \Gamma}_m$. Equation (\ref{eq:emme}) implies ${{\bf{M}}_m} \cdot {{\bf{Y}}_m} =
{{\bf{\Gamma }}_m}$. Let the subscript "c" label likewise the same quantities as
estimated by transport codes solving Eq. (\ref{eq:fpe}) via least squares. Finally, let
"*" label the "true" (unknown) quantities that the experimental measure is
addressing:   ${\bf M}_*, \zeta_*, {\bf \Gamma}_*$, and  ${{\bf{M}}_*} \cdot {{\bf{Y}}_*} =
{{\bf{\Gamma }}_*}$. ${\bf \Gamma}_m$  generally contains some arbitrariness due to the lack of
precise knowledge about the source term, too.
Let $\delta \zeta  = {\zeta _c} - {\zeta _m}\,,\;\delta {\bf{M}} = {{\bf{M}}_c} -
{{\bf{M}}_m}\,,\;{\delta _*}{\bf{Y}} = {{\bf{Y}}_c} - {{\bf{Y}}_*}$ . The target of any
transport modeling is ${\delta _*}{\bf{Y}} = 0$. Finally, we write ${{\bf \Gamma} _c} = {\bf{\Gamma}
_*} + \Delta {{\bf \Gamma}_S} + \Delta {{\bf \Gamma}_m} + \Delta {{\bf \Gamma}_\zeta }$. In this
expression $\Delta {{\bf \Gamma}_S}$ is the error in the calculation of the flux due to the imperfect
knowledge of the source, $\Delta {{\bf \Gamma}_m}$  is the error related to imprecise measurement
of $\zeta$: ${\zeta _m} - {\zeta _*}$, and the third term comes from the error in the
reconstruction of the measured $\zeta$: $\delta \zeta$. For brevity, we set
${{\bf \Gamma}_\Delta } = {{\bf \Gamma}_*} + \Delta {{\bf \Gamma}_S} + \Delta {{\bf \Gamma}_m}\,,\;
{{\bf \Gamma}_c} = {{\bf \Gamma}_\Delta } + \Delta {{\bf \Gamma}_\zeta }$.
Since (\ref{eq:emme}) is a first integral of (\ref{eq:fpe}), ${{\bf{M}}_c} \cdot {{\bf{Y}}_c} =
{{\bf{\Gamma }}_c}$  holds. Using above definitions, we rewrite the latter expression as
\begin{equation}
{{\bf{M}}_{\bf{m}}} \cdot {{\bf{\delta }}_*}{\bf{Y = }}{{\bf{\Gamma }}_{\bf{\Delta
}}}{\bf{ - }}{{\bf{M}}_{\bf{m}}} \cdot {{\bf{Y}}_{\bf{*}}}{\bf{ + \Delta
}}{{\bf{\Gamma }}_{\bf{\zeta }}}{\bf{ - \delta M}} \cdot {{\bf{Y}}_{\bf{*}}}{\bf{ - \delta
M}} \cdot {{\bf{\delta }}_*}{\bf{Y}}
\label{eq:emmem}
\end{equation}
Whenever $\delta \zeta  = {\zeta _c} - {\zeta _m}\, = 0$, then $\;\delta {\bf{M}} = {{\bf{M}}_c} -
{{\bf{M}}_m}\, = 0,\,\Delta {{\bf \Gamma}_\zeta} = 0$. Let again $\lambda_0$ be the smaller
eigenvalue (in absolute value). If $\lambda_0 \neq 0$ it is possible to check that
(\ref{eq:emmem}) can be fulfilled by setting $\delta \zeta =0$: for regular points, transport codes can
potentially attain a perfect reconstruction of measured value, provided the set of trial profiles includes the solution given by the matricial method. 
\\
This is no longer true when
$\lambda_0 = 0$, since in this case the l.h.s. of (\ref{eq:emmem}) is aligned along the other
eigenvector ${\bf E}_1$, whereas the r.h.s. has generically components along both directions.
At a singular point a transport code provides a natural regularization because of the limited
flexibility of the set of $\bf Y$ profiles available for the minimization. In terms of Eq.
(\ref{eq:emmem}) the code generates the minimum finite value of $\delta \zeta(r_s)$ over the set
of trial profiles providing a finite value of $\delta {\bf M}_\zeta(r_s)$ which cancels the ${\bf
E}_0$ component of $\delta {\bf \Gamma}(r_s)$, that exists for $\delta \zeta(r_s) =0$. More precisely the choice of all the $\delta \zeta(r_s)$'s is done simultaneously, because of the global norm used by the codes. Therefore this optimized
finite value of $\delta \zeta(r_s)$ may not be the smallest one. It is independent of $\delta_* {\bf Y}(r_s)$ if $\delta {\bf M} \, \delta
\zeta(r_s)$ is small. Therefore $\delta \zeta(r_s)$ does not measure the accuracy of the calculation
of ${\bf Y}(r_s)$. Since $\delta {\bf M} \, \delta \zeta(r_s)$  has a random sign, it does not improve
this accuracy, even if it is not negligible. Finally the finite value of $\delta \zeta(r_s)$
also modifies the component of ${\bf Y}(r_s)$ along ${\bf E}_1$, which increases the
uncertainty of the calculation. \\
Summarizing, the matricial approach: (I) highlights that the source term plays a critical role for
profile calculation; the best situation is obtained when the source is localized at the plasma outer
edge: then, the inverse problem is well-conditioned and the error on the source estimate does not
enter the calculation of the profile for smaller radii. The worst case is conversely expected to be when the source
is spred radially. Then the calculated profile may be strongly influenced by the somewhat
arbitrary regularization. (II) Points out that the common strategy of least squares minimization
performed by transport codes actually hides some subtle practicalities. \\
So far, we have implicitly assumed that measurements are performed on a very fine mesh of
points, such that $\zeta$ may be treated as a continuous variable. In real cases, measurements are
taken on a discrete and often quite coarse mesh, $ r = r_j, j = 1,...,N$. This raises issues of
interpolation and extrapolation, needed to compute the derivatives and the integrals involved in
Eq. (\ref{eq:emme}), and emphasizes
the intrinsic reverse character of our approach with respect to transport codes solving Eq.
(\ref{eq:fpe}): the matricial approach computes $\bf Y$ only at the discrete set of measurement
points $r_j$, but needs a continuous interpolation of the data $A$ and $\phi$ in order to compute
the derivatives involved in Eq. (\ref{eq:emme}). Conversely, transport codes do need continuous
profiles for $(\chi, V)$ and only the knowledge of $A, \phi$ at discrete points. \\
From Eq. (\ref{eq:y}) one can estimate the uncertainty associated to the calculation of ${\bf
Y}(r)$.  $\bf Y$ depends on $3\times N$ experimental quantities $\left\{ {{A_i}} \right\},\left\{
{{\phi _i}} \right\},\left\{ {{S_i}} \right\}$. We label collectively these quantities as ${\vartheta
_k},\,k = 1,...,3N$. Let $\delta {\vartheta _k}$  be an estimate of the associated uncertainty
and
$ {{\partial _k} \equiv \frac{\partial }{{\partial {\vartheta _k}}}}$. Thus:
\begin{equation}
\begin{array}{l}
 \delta {\bf{Y}} = \sum\limits_{k,l} {\delta {\vartheta _k}} \left[ {\frac{{{\bf{\Gamma }} \cdot
{{\bf{E}}_l}}}{{{\lambda _l}}}{\partial _k}{{\bf{E}}_l}} \right. \\
 \begin{array}{*{20}{c}}
   {\begin{array}{*{20}{c}}
   {} & {}  \\
\end{array}} & {} & {\left. { + \,{{\bf{E}}_l}\left( {\frac{{{\partial _k}{\bf{\Gamma }} \cdot
{{\bf{E}}_l}}}{{{\lambda _l}}} + \frac{{{\bf{\Gamma }} \cdot {\partial
_k}{{\bf{E}}_l}}}{{{\lambda _l}}} - \frac{{{\bf{\Gamma }} \cdot {{\bf{E}}_l}}}{{\lambda
_l^2}}{\partial _k}{\lambda _l}} \right)} \right]}  \\
\end{array} \\
 \end{array}
\label{eq:deltay}
\end{equation}
Generally, the  $\delta {\vartheta _k}$'s should be treated as stochastic variables picked up
from some statistical distribution. Accordingly the vector $\delta {\bf{Y}}(r_j)$ spans an area
$W$ around the point ${\bf Y}(r_j)$: the uncertainty on this quantity is geometrically displayed
as a roughly elliptic region with low eccentricity if ${\lambda _0}/{\lambda _1} \approx O(1)$ .
Near the points where $\lambda_0$ is small the term proportional to $\lambda _0^{ - 2}$
dominates:
\begin{equation}
\delta {\bf{Y}} \approx  - {{\bf{E}}_0}\frac{{{\bf{\Gamma }} \cdot
{{\bf{E}}_0}}}{{\lambda _0^2}}\sum\limits_k {d{\vartheta _k}} \left( {{\partial _k}{\lambda
_0}} \right) + {{\bf{E}}_1} O\left( {\frac{1}{{{\lambda _0}}}} \right)
\end{equation}
Thus $W$ is stretched along ${\bf E}_0$, as qualitatively assessed earlier. 

Till now we considered a single experimental setup endowed with a single modulation frequency
and source. However it is easy to generalize
to experiments where--under the same transport conditions--multiple harmonics are measured, or the
location of sources is changed. Then, the above procedure may be repeated, and the independent
domains of uncertainty $W$ compared. If for all points they have a non vanishing intersection,
this should provide a smaller global uncertainty, and thus improve the accuracy of the profile
calculation. If one of the intersections vanishes, the assumptions of the calculation must be
questioned. In particular one may wonder whether: (I) the uncertainties  have been
underestimated and whether the correct $W$'s are larger than estimated; (II) Some source
terms have not been computed correctly or are missing; (III) It is not true that the independent
measurements correspond to the same $(\chi(r),V(r))$ profile; (IV) Eq. (\ref{eq:fpe}) fails:
transport is not of the advection-diffusion type.
Whenever the modulated source has several harmonic components, by virtue of the linearity of
Eq. (\ref{eq:fpew}), each harmonic can be treated as a separate measurement. This may improve
the accuracy in the regions where $\lambda_0$ is not small. If $\lambda_0$ is small, since ${\bf
E}_0$ has almost the same orientation for all harmonics, the precision of the calculation cannot be
improved. \\
In order to make visual the above statements we propose here below a check using synthetic data.
Spatial profiles of transport coefficients and
sources are given in advance (Fig. 1a,b) and used in Eq. (\ref{eq:fpew}) to produce synthetic
$(A,\phi)$ datasets (Fig. 1c,d). Notice that the central source $S_2$ produces a singular point close to half radius (i.e., near the location of its peak), whereas the edge source $S_1$ does not. The matricial method is then employed on these data to check that
the original transport coefficients are correctly recovered back (Fig. 1e,f). Finally, we add
 a finite uncertainty to the "measurements" in order to mimic experimental error: to each point is added a
random contribution taken from a normal distribution with a mean amplitude equal to $1\%$
(both in the amplitude and the phase). The new datasets have been then smoothed using a moving
average, in order to avoid too brusque variations, that would deteriorate the quality of the
derivatives, and were interpolated using 3rd-order splines. Then the coefficients $(\chi, V)$ are recomputed, repeating the
whole procedure for a total of 400 independent statistical runs at $x = 0.41$, i.e., close to the singularity for the source $S_2$. Figs. (1g,h) show how the
different estimates spread around the true value. For the "regular" case $S_1$ all the estimates have a relatively small spread, unlike $S_2$, whose data align along the eigenvector ${\bf
E}_0$, as predicted earlier, when $\lambda_0 \to 0$. The width of the spread is remarkable,
which stresses again the ill-posed nature of the problem. 
\begin{figure}
\includegraphics[width=70mm]{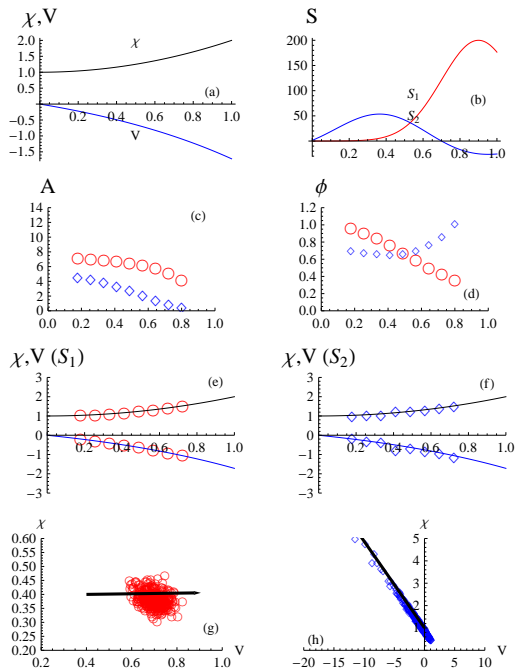}
\caption{(a) Profiles of $\chi$ and $V$; (b) Profiles of the two sources, $S_1, S_2$.
Both sources are modulated with frequency
$\nu = 1$. (c) Amplitude $A$ of the signal produced by solving Eq. (\ref{eq:fpew}). Circles
refer to source $S_1$, diamonds to $S_2$. (d) Phase $\phi$ of the same signal. (e)
Symbols are the transport coefficients computed backwards from Eq. (\ref{eq:emme}),
superimposed to the true transport coefficients,
for the simulation with source $S_1$. (f) The same for source $S_2$. (g) Scatter of $(\chi, V)$ from MonteCarlo simulation using source $S_1$ at $x = 0.41$. The black segment is parallel
to the local eigenvector $E_0$. (h) The same as  figure (g) for data from source $S_2$. }
\label{fig:figura}
\end{figure}

To summarize, the matricial approach (MA) is very light computationally. Indeed it avoids the heavy spatio-temporal integration of Eqns.
(\ref{eq:fpe},\ref{eq:fpew}) and the iterative least-squares minimization procedure over the set of
trial functions. The MA provides a clear geometrical foundation to the nature and size of uncertainties in profile
reconstruction. The reconstruction radius-by-radius enables to see how different are the
uncertainties over $\bf Y$ as a function of $r$ and to correlate them with the presence of a
source. The MA yields a higher precision in the reconstruction of transport profiles than transport codes,
provided the uncertainty on the estimate of the derivatives of $A$ and $\phi$ is not high. Indeed
the MA is not restricted by the a-priori guess of the trial profiles, but by that of these
derivatives, which is much more reliable and controllable. It may provide an assessment of profile
predictions already done with transport codes. A posteriori some regularization may be applied to MA results. 
When $\lambda_0$ is small at a given radius $r_s$, the regularization needed to provide a
reasonable estimate of ${\bf Y}(r_s)$ can be defined in an explicit way, while this is only implicit
in the transport code approach.
For instance, if a singular point $r_s$ is in
between two nearby regular ones $r_i$ and $r_{i+1}$, one may require ${\bf Y}(r_s)$ to be on the
straight line joining ${\bf Y}(r_i)$ and ${\bf Y}(r_{i+1})$. Overlapping the uncertainty intervals for various experiments where the same transport is
assumed to hold, provides either a way to improve the precision of the reconstruction (case of a
non vanishing overlap) or to prove the set of initial assumptions in the reconstruction to be wrong
(case of a vanishing overlap). The MA can help designing a priori the combination of perturbation measurements susceptible of
improving the precision of the reconstruction of transport profiles. The MA requires a single boundary condition only: the vanishing flux at $r=0$, which is a rigorous constraint. In contrast the calculation via Fokker-Planck equation of the $\zeta_i(t)$'s requires a second outer boundary condition which is generally known with some uncertainty. Very often this condition is provided by the data of the outermost chord. Then the matricial approach has the benefit to keep the outermost chord data available for the profile calculation. It also avoids the uncertainty included in this data to contaminate the profile calculation at smaller radii.

This work resulted from a long series of discussions with P. Mantica about analysis of JET data. We thank V. F. Andreev,  Y. Camenen, X. Garbet, and A. Salmi for remarks about a preliminary discussion paper which drove us into clarifying several issues, R. Paccagnella for very useful suggestions to improve the clarity of the contents, and Y. Elskens for useful suggestions about the wording. This work was supported by EURATOM and carried out within
the framework of the European Fusion Development
Agreement. The views and opinions expressed herein do not necessarily reflect those of the
European Commission.

\end{document}